\documentclass[aps,prb,twocolumn,superscriptaddress,longbibliography]{revtex4-2}
\usepackage[pdftex]{graphicx}
\usepackage{mmap}
\usepackage{amssymb}
\usepackage{amsmath}
\usepackage{color}
\usepackage[usenames,dvipsnames]{xcolor}
\usepackage{hyperref}
\hypersetup{colorlinks=true,linkcolor=blue,urlcolor=blue,citecolor=blue}

\begin{document}
\title{Drift velocity of edge magnetoplasmons due to magnetic edge channels}
\author{Alexey A. Sokolik}
\email{asokolik@hse.ru}
\affiliation{Institute for Spectroscopy RAS, 108840 Troitsk, Moscow, Russia}
\affiliation{Institute of Microelectronics Technology and High Purity Materials, Russian Academy of Sciences, Chernogolovka 142432, Russia}
\affiliation{National Research University Higher School of Economics, 109028 Moscow, Russia}
\author{Yurii~E. Lozovik}
\affiliation{Institute for Spectroscopy RAS, 108840 Troitsk, Moscow, Russia}
\affiliation{Institute of Microelectronics Technology and High Purity Materials, Russian Academy of Sciences, Chernogolovka 142432, Russia}
\affiliation{National Research University Higher School of Economics, 109028 Moscow, Russia}

\begin{abstract}
Edge magnetoplasmons arise on a boundary of conducting layer in perpendicular magnetic field due to an interplay of electron cyclotron motion and Coulomb repulsion. Lateral electric field, which confines electrons inside the sample, drives their spiraling motion in magnetic field along the edge with the average drift velocity contributing to the total magnetoplasmon velocity. We revisit this classical picture by developing fully quantum theory of drift velocity starting from analysis of magnetic edge channels and their electrodynamic response. We derive the quantum-mechanical expression for the drift velocity, which arises in our theory as a characteristic of such response and can be calculated as harmonic mean of group velocities of edge channels crossing the Fermi level. Using the Wiener-Hopf method to solve analytically the edge mode electrodynamic problem, we demonstrate that the edge channel response effectively enhances the bulk Hall response of the conducting layer and thus increases the edge magnetoplasmon velocity. In the long-wavelength limit of our model, the drift velocity is simply added to the total magnetoplasmon velocity, in agreement with the classical picture.
\end{abstract}

\maketitle

\section{Introduction}
Edge magnetoplasmons on a boundary of two-dimensional electron gas (2DEG) are induced jointly by electric field from electron density perturbations near the edge and by Lorentz force from the external magnetic field \cite{Fetter}. Due to the time-reversal symmetry breaking in magnetic field, edge magnetoplasmons propagate along the sample boundary only in one direction and cannot scatter back. This property, together with strong confinement of plasma oscillations near the boundary, is essential to possible applications of edge magnetoplasmons for energy and signal transmission on micro- and nanoscale \cite{Hiyama,Hashisaka,Jin,Mahoney}. Edge magnetoplasmons were studied experimentally on boundaries of 2DEG formed by depleting gate electrodes \cite{Wassermeier,Ashoori,Talyanskii1,Ernst,Mast,Talyanskii2,Kamata,Kumada1,Andreev,Endo,Wu,Kumada2}, on graphene edges \cite{Crassee,Yan,Kumada3,Kumada4,Petkovic1,Lin,Petkovic2,Kumada5}, and, recently, on the edges of quantum anomalous Hall insulators which do not require external magnetic field to support magnetoplasmons \cite{Mahoney2017,Wang2023,Martinez}.

\begin{figure}[t]
\begin{center}
\includegraphics[width=0.72\columnwidth]{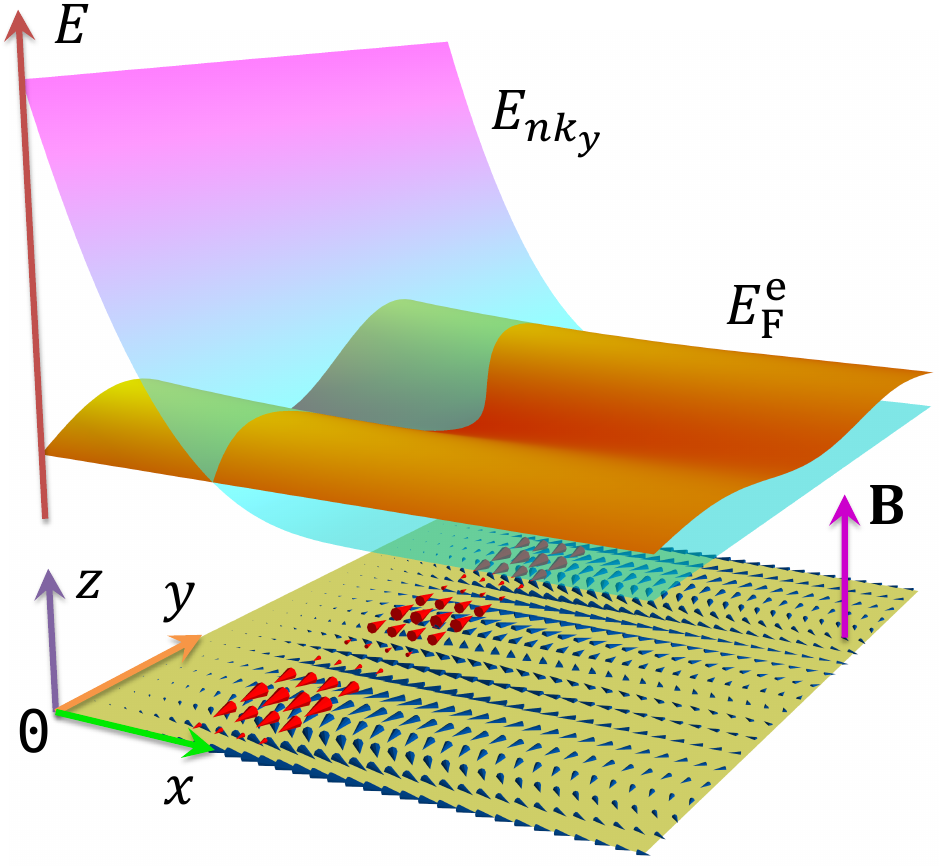}
\end{center}
\caption{\label{Fig_picture}Illustration of the origin of edge magnetoplasmon drift velocity. Wavelike oscillations of the local Fermi level $E_\mathrm{F}^\mathrm{e}$ change the filling of magnetic edge channels, i.e. Landau levels with the energies $E_{nk_y}$, which are bent upwards near the edge $x=0$. Since the edge channels are unidirectional, oscillations of their occupation give rise to oscillations of edge current shown by red cones. The edge current appears in phase with the bulk current response of the 2DEG (blue cones) and thus effectively enhances its Hall conductivity $\sigma_{xy}$. The edge magnetoplasmon velocity, which is roughly proportional to $|\sigma_{xy}|$, is increased too by the quantity $v_\mathrm{dr}$ which has the meaning of drift velocity.}
\end{figure}

Theoretical treatment of edge magnetoplasmons relies on various approximations applied to solve Maxwell equations for a conducting half-plane \cite{Fetter,Wang,Aleiner,Johnson} or on the exact Wiener-Hopf method which allows to solve the equations analytically \cite{Volkov,Margetis,Sokolik}. Additional factors such as dissipation \cite{Talyanskii2,Kumada3,Johnson,Sokolik} and formation of alternating strips of compressible and incompressible phases near the edge \cite{Talyanskii1,Kumada1,Andreev,Endo,Johnson,Aleiner,Chklovskii,Balev1,Balev2,Sukhodub} make the theory of edge magnetoplasmons more complicated. The basic feature of a local structure of electronic states near the edge in magnetic field is formation of magnetic edge channels \cite{Halperin} due to the upward bending of Landau levels caused either by hard boundary of the sample (see Fig.~\ref{Fig_picture}) or by soft confinement field. In the latter picture with the soft edge taken at the classical level, the lateral electric field which confines electrons inside a sample drives electrons into spiraling motion in crossed electric and magnetic fields with some average \emph{drift velocity}. Switching to the reference frame moving with the drift velocity, where the electric field vanishes, we obtain simple addition of the drift velocity to those magnetoplasmon velocity which is predicted by the edge mode theory without taking into account the carrier drift. Such classical-level addition of velocities was used to interpret data of several experiments \cite{Kumada4,Petkovic1,Martinez}. Alternative explanations of the extra edge magnetoplasmon velocity are based on local capacitance models \cite{Wassermeier,Talyanskii1,Endo}, and the recent analysis uses the model of chiral  linearly-dispersive edge states \cite{Wang2023}.

In this paper, we develop the fully quantum theory of drift velocity of edge magnetoplasmons, which starts from the analysis of electron states near the hard edge of 2DEG in magnetic field (magnetic edge channels), presented in Sec.~\ref{Sec_Edge}. As shown schematically in Fig.~\ref{Fig_picture}, perturbations of a local Fermi level near the edge during magnetoplasma oscillations induce the linear responses of oscillating charge and current densities, which are also confined to the edge on a scale of magnetic length. The drift velocity $v_\mathrm{dr}$ arises in our theory as the joint characteristic of charge and density responses of the magnetic edge channels, and we derive the analytical expression for it as harmonic mean of group velocities of those edge states which cross the Fermi level. The additional current and density responses of the edge channels, which should be taken into account together with the bulk response of 2DEG, give rise to additional terms in equations describing the edge magnetoplasmons. Solving these equations analytically using the Wiener-Hopf method in Sec.~\ref{Sec_WH}, we show that the edge response effectively enhances the bulk Hall conductivity of 2DEG and thus increases the total magnetoplasmon velocity. In Sec.~\ref{Sec_calc} we present calculations of the edge magnetoplasmon dispersion and analyze its long-wavelength limit. We show that in this limit $v_\mathrm{dr}$ is approximately added to the magnetoplasmon velocity obtained without taking into account the edge response. Moreover, in the limit of large number of occupied Landau levels, $v_\mathrm{dr}$ tends to the average velocity of electron spiraling motion in the effective confining electric field superimposed on the magnetic field. These results agree with the conventional classical picture, but here we obtain them from the first principles. Sec.~\ref{Sec_concl} is devoted to conclusions, and Appendix~\ref{Appendix_A} provides details about quasiclassical calculations of the drift velocity.

\section{Edge channels and their response}\label{Sec_Edge}

Consider the edge states of electrons on the boundary of 2DEG, which occupies the half-plane $x\geqslant0$, $z=0$. The uniform external magnetic field $\mathbf{B}=\mathbf{e}_zB$ corresponds to the vector potential $\mathbf{A}=\mathbf{e}_yBx$ taken in the Landau gauge. Electron wave functions can be found as $\psi(x,y)=L_y^{-1/2}e^{ik_yy}\phi(x)$, where $k_y$ is the electron momentum along the edge, $L_y$ is the system size in the $y$ direction, and $\phi(x)$ obeys the Schrodinger equation
\begin{equation}
-\frac{\hbar^2}{2m}\phi''(x)+\frac1{2m}\left(\hbar k_y+\frac{eBx}c\right)^2\phi(x)=E\phi(x)\label{Schrodinger}
\end{equation}
(hereafter we take the electron charge equal to $-e$, where $e>0$). For a hard edge, we impose the Dirichlet boundary condition $\phi(0)=0$. General solution of Eq.~(\ref{Schrodinger}), which is normalizable at the half-line $x\geqslant0$, is given by the parabolic cylinder functions $U(-\varepsilon,z)$ \cite{Dean}. The full electron wave function reads
\begin{equation}
\psi_{nk_y}(x,y)=\frac{e^{ik_yy}}{\sqrt{L_yl_H}N_{nk_y}}U\left(-\varepsilon_{nk_y},\frac{\sqrt{2}}{l_H}(x+l_H^2k_y)\right),\label{wf_edge}
\end{equation}
and the corresponding energy of the $n$th stationary state $E_{nk_y}=\hbar\omega_\mathrm{c}\varepsilon_{nk_y}$ is related to the dimensionless energy $\varepsilon_{nk_y}$ which can be found as the $(n+1)$th (in the increasing order, $n=0,1,2,\ldots$) root of the equation
\begin{equation}
U(-\varepsilon_{nk_y},\sqrt{2}l_Hk_y)=0.\label{energy_equation}
\end{equation}
Here $\hbar\omega_\mathrm{c}=\hbar eB/mc$ and $l_H=\sqrt{\hbar c/eB}$ are, respectively, cyclotron energy and magnetic length, which determine characteristic energy and length scales of electron quantum states in magnetic field. The factor $N_{nk_y}$ is determined by the wave function normalization condition $\int_0^\infty dx\int_0^{L_y}dy\,|\psi_{nk_y}(x,y)|^2=1$.

\begin{figure}[t]
\begin{center}
\includegraphics[width=0.75\columnwidth]{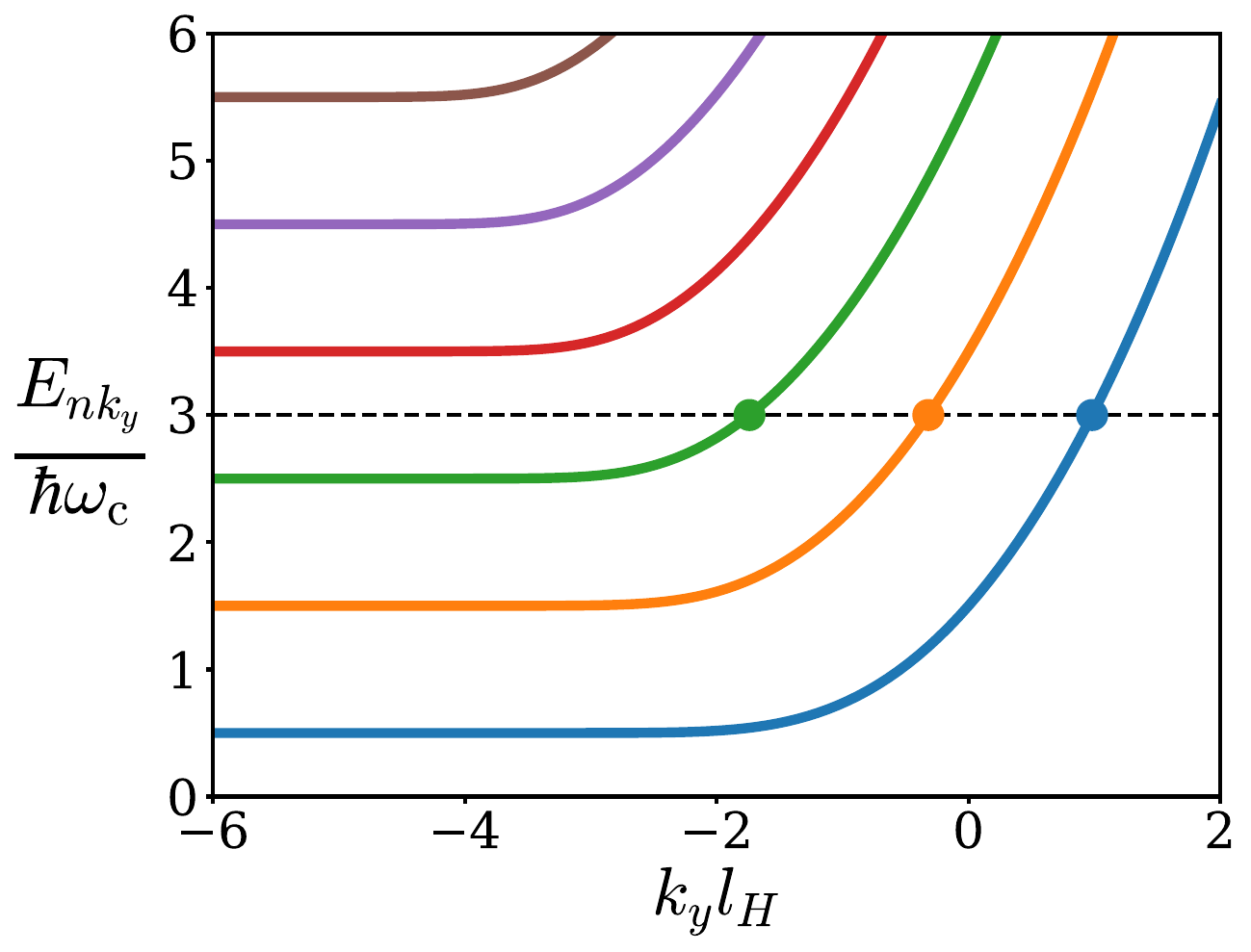}
\end{center}
\caption{\label{Fig_energies} Energies of the edge states $E_{nk_y}$ in magnetic field as functions of the wave vector $k_y$ along the edge. The curves from bottom to top correspond to the levels $n=0\ldots5$. Dashed line show the example of Fermi level location $E_\mathrm{F}^\mathrm{e}$ between the bulk Landau levels $n=2$ and $3$, and circles demonstrate its intersections with the edge state energies at momenta $k_n^\mathrm{max}$, $n=0,1,2$.}
\end{figure}

\begin{figure}[t]
\begin{center}
\includegraphics[width=\columnwidth]{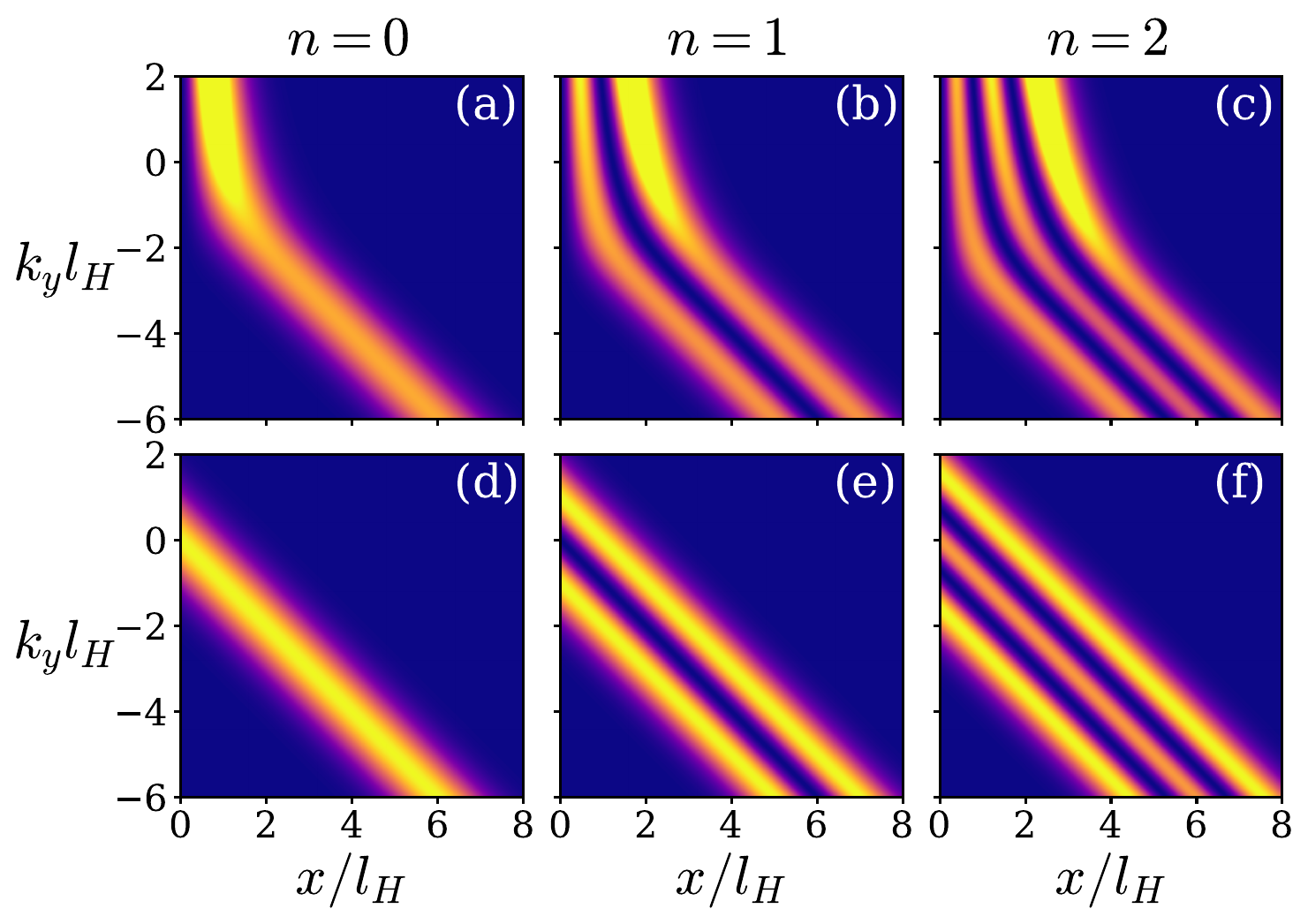}
\end{center}
\caption{\label{Fig_wf}Wave function square moduli $|\psi|^2$ for (a)--(c) edge states (\ref{wf_edge}), (d)--(f) bulk Landau levels (\ref{wf_Landau}) at different wave vectors $k_y$ along the edge. The columns correspond to the level numbers $n=0,1,2$.}
\end{figure}

Edge state energies at different $n$ as functions of $k_y$ are shown in Fig.~\ref{Fig_energies}. At $k_y\rightarrow-\infty$ the center of wave function accumulation $x\approx-l_H^2k_y$ moves away from the edge $x=0$ to the bulk $x\rightarrow\infty$. The influence of the boundary weakens, so the edge state wave functions and energies tend to those of ordinary Landau levels \cite{Landau} in unbounded 2DEG in magnetic field:
\begin{align}
\psi_{nk_y}^\mathrm{L}(x,y)&=\frac{e^{ik_yy}e^{-(x+l_H^2k_y)^2/2l_H^2}}{\sqrt{2^nn!L_yl_H\sqrt\pi}}H_n\left(\frac{x+l_H^2k_y}{l_H}\right),\label{wf_Landau}\\
E_n^\mathrm{L}&=\hbar\omega_\mathrm{c}\left(n+\frac12\right),
\end{align}
where $H_n$ are Hermite polynomials. In the opposite limit $k_y\rightarrow\infty$, the wave functions are strongly squeezed against the edge potential wall, so the energies tend to the classical limit $E_{nk_y}\sim\hbar^2k_y^2/2m$, which describes electron states freely propagating along the edge. Appendix~\ref{Appendix_A} presents more accurate formula (\ref{E_qc_appr}) for edge state energies applicable at $l_Hk_y>-\sqrt{2n+1}$, which is derived using quasiclassical quantization. Fig.~\ref{Fig_wf}(a)--(c) shows square moduli of the edge state wave functions (\ref{wf_edge}) for $n=0,1,2$, and Fig.~\ref{Fig_wf}(d)--(f) shows square moduli of bulk Landau level (or displaced harmonic oscillator) wave functions (\ref{wf_Landau}) which could exist without the Dirichlet boundary condition at $x=0$. The wave functions in the top and bottom rows become progressively similar at $k_y\rightarrow-\infty$.

Occupation of the edge states (\ref{wf_edge}) by electrons gives rise to the following spatial profiles of charge density and current density along the edge:
\begin{align}
\rho_\mathrm{e}(x)&=-eg\sum_{nk_y}f_{nk_y}|\psi_{nk_y}(x,y)|^2,\label{rho_e_gen}\\
j_y(x)&=-\frac{eg}m\sum_{nk_y}f_{nk_y}\left(\hbar k_y+\frac{eBx}{c}\right)|\psi_{nk_y}(x,y)|^2.\label{j_e_gen}
\end{align}
Here $g=2$ is the spin degeneracy factor for electron states, $f_{nk_y}$ are occupation numbers of edge states which, at low temperatures, follow the step-like energy dependence $f_{nk_y}=\Theta(E_\mathrm{F}^\mathrm{e}-E_{nk_y})$, where $E_\mathrm{F}^\mathrm{e}$ is the local Fermi level near the edge, and $\Theta(x)$ is the unit step function. We need to sum over momenta $k_y$ of the occupied edge states in Eqs.~(\ref{rho_e_gen})--(\ref{j_e_gen}), which are integer multiples of $2\pi/L_y$ in a sample with periodic boundary conditions along the $y$ axis. As seen from Fig.~\ref{Fig_energies}, this summation should be carried out from $-\infty$ to the maximal values $k_n^\mathrm{max}$, which are different for each $n$. These values are given by the condition $E_{nk_n^\mathrm{max}}=E_\mathrm{F}^\mathrm{e}$ or, using Eq.~(\ref{energy_equation}), by the equation \begin{equation}
U(-E_\mathrm{F}^\mathrm{e}/\hbar\omega_\mathrm{c},\sqrt2l_Hk_n^\mathrm{max})=0.
\end{equation}
Separating in Eqs.~(\ref{rho_e_gen})--(\ref{j_e_gen}) contributions of different levels $n$ as $\rho_\mathrm{e}(x)=\sum_n\rho_{\mathrm{e}n}(x)$, $j_y(x)=\sum_nj_{yn}(x)$, and switching from summation over $k_y$ to integration in the limit $L_y\rightarrow\infty$, we obtain
\begin{align}
\rho_{\mathrm{e}n}(x)=-\frac{eg}{2\pi l_H}\int\limits_{-\infty}^{k_n^\mathrm{max}}dk_y\frac{U^2(-\varepsilon_{nk_y},\sqrt2(k_yl_H+x/l_H))}{|N_{nk_y}|^2},\label{rho_e_n}
\end{align}
\begin{align}
j_{yn}(x)=-\frac{eg\omega_c}{2\pi}&\int\limits_{-\infty}^{k_n^\mathrm{max}}dk_y(k_yl_H+x/l_H)\nonumber\\ &\times\frac{U^2(-\varepsilon_{nk_y},\sqrt2(k_yl_H+x/l_H))}{|N_{nk_y}|^2}.\label{j_e_n}
\end{align}
Note that $\rho_{\mathrm{e}n}(x)=0$, $j_{yn}(x)=0$ when $E_n^\mathrm{L}\leqslant E_\mathrm{F}^\mathrm{e}$, because in this case the whole $n$th edge state is unoccupied.

\begin{figure}[t]
\begin{center}
\includegraphics[width=\columnwidth]{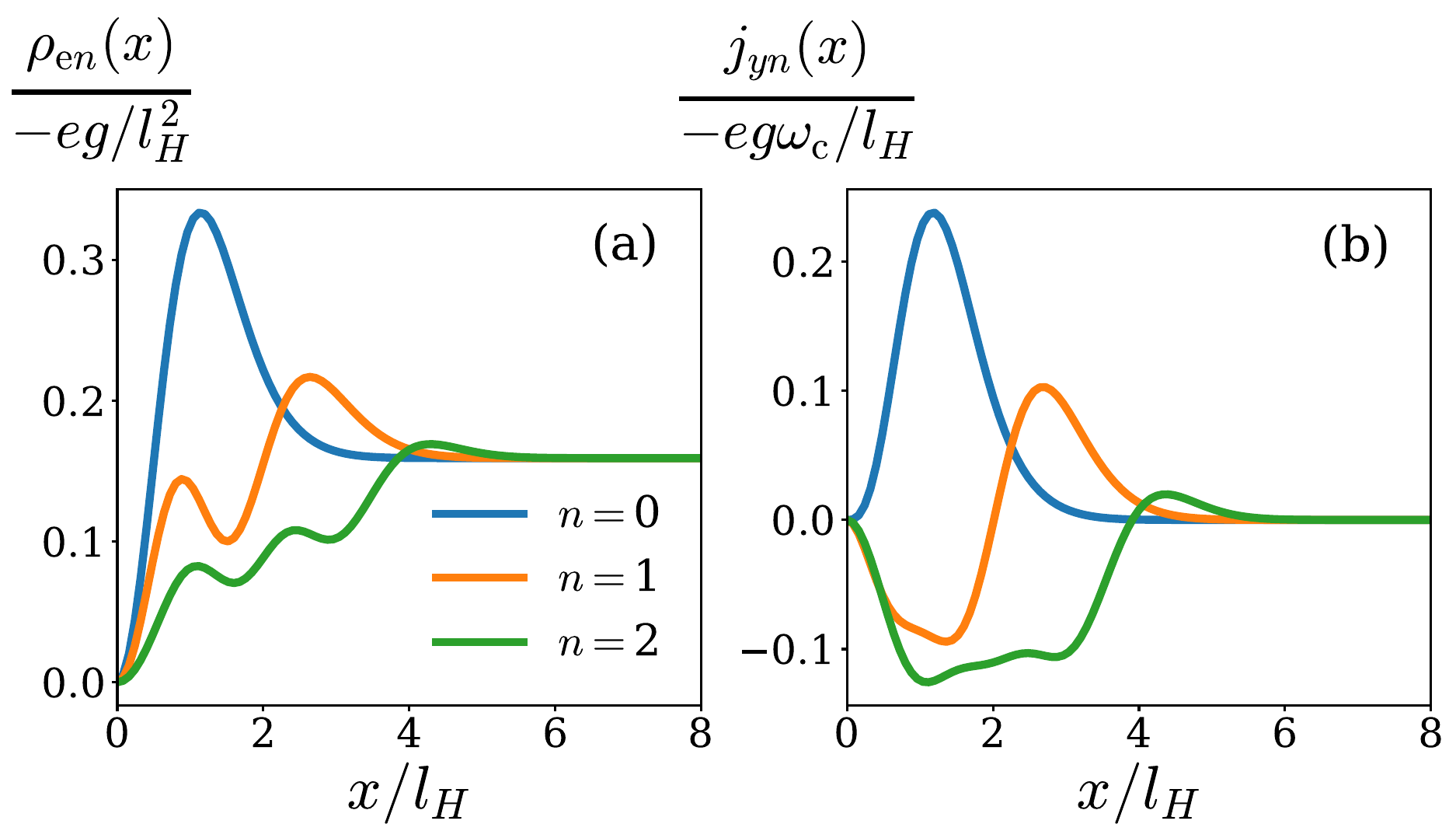}
\end{center}
\caption{\label{Fig_dens_curr}Contributions (\ref{rho_e_n})--(\ref{j_e_n}) of individual edge channels $n=0,1,2$ to (a) charge density $\rho_\mathrm{e}(x)$ and (b) current density $j_y(x)$ as functions of the distance $x$ from the edge. The Fermi level $E_\mathrm{F}^\mathrm{e}$ is located in the mid-gap between the bulk Landau levels $n=2$ and 3, as shown in Fig.~\ref{Fig_energies}.}
\end{figure}

Fig.~\ref{Fig_dens_curr} shows the example of how the $n$th edge channels contribute to spatial profiles of the charge and current densities. These profiles are confined to the distance $x\sim l_H$ of several magnetic lengths from the edge $x=0$. Note that each charge density profile $\rho_{\mathrm{e}n}(x)$ tends at $x\rightarrow\infty$ to the universal value $-eg/2\pi l_H^2$ equal to the charge density of a single completely filled bulk Landau level. Since we are interested only in variations of charge density during magnetoplasma oscillations due to transient changes of edge state occupations, we subtract this constant part by considering not the charge density itself $\rho_\mathrm{e}(x)$, but its local deviation from the bulk value near the edge: $\tilde\rho_{\mathrm{e}n}(x)=\rho_{\mathrm{e}n}(x)+eg/2\pi l_H^2$, $\tilde\rho_\mathrm{e}(x)=\rho_\mathrm{e}(x)+n_0eg/2\pi l_H^2$, where $n_0$ is the number of completely filled bulk Landau levels, which depends on the bulk Fermi level location.

Taking into account that local variations of charge and current densities near the edge are confined to a region of several magnetic lengths, which is small in strong magnetic field in the quantum Hall regime, we can look only on the integral charge (per unit length of the edge) $Q_\mathrm{e}=\sum_nQ_{\mathrm{e}n}$, $Q_{\mathrm{e}n}=\int_0^\infty dx\,\tilde\rho_{\mathrm{e}n}(x)$, and integral current along the edge $I_y=\sum_nI_{yn}$, $I_{yn}=\int_0^\infty dx\,j_{yn}(x)$ both resolved over edge channels numbers $n$. These quantities can be found analytically. First, representing the bulk charge density $-eg/2\pi l_H^2$ as the $k_y$-integrated density $-eg|\psi_{nk_y}^\mathrm{L}|^2$ provided by the bulk Landau levels (\ref{wf_Landau}), we obtain from Eq.~(\ref{rho_e_n}):
\begin{align}
Q_{\mathrm{e}n}=&-\frac{egL_y}{2\pi}\int\limits_0^\infty dx\left\{\int\limits_{-\infty}^{k_n^\mathrm{max}}dk_y\:|\psi_{nk_y}(x,y)|^2\right.\nonumber\\
&\left.-\int\limits_{-\infty}^{\infty}dk_y\:|\psi_{nk_y}^\mathrm{L}(x,y)|^2\vphantom{\int\limits_{-\infty}^{k_n^\mathrm{max}}}\right\}=-\frac{eg}{2\pi}k_n^\mathrm{max}.\label{Q_en}
\end{align}
The last equality was obtained by changing the order of integration over $x$ and $k_y$ (which is possible due to rapid decay of the integrand $|\psi_{nk_y}(x,y)|^2-|\psi_{nk_y}^\mathrm{L}(x,y)|^2$ at ${x\rightarrow\infty}$, $k_y\rightarrow-\infty$, see Fig.~\ref{Fig_wf}) and using the symmetry property $|\psi_{nk_y}^\mathrm{L}(x,y)|^2=|\psi_{n,-k_y}^\mathrm{L}(-x,y)|^2$. Second, using the Hellmann-Feynman theorem $\partial E_{nk_y}/\partial k_y=\langle\psi_{nk_y}|\partial H/\partial k_y|\psi_{nk_y}\rangle$, we can easily integrate Eq.~(\ref{j_e_n}) by $x$, so the integral edge current reads
\begin{equation}
I_{yn}=-\frac{eg}{2\pi\hbar}\int\limits_{-\infty}^{k_n^\mathrm{max}}dk_y\:\frac{\partial E_{nk_y}}{\partial k_y}=-\frac{eg}{2\pi\hbar}(E_\mathrm{F}^\mathrm{e}-E_n^\mathrm{L}).\label{I_yn}
\end{equation}
Summing contributions of all partially occupied edge channels with $E_\mathrm{F}^\mathrm{e}>E_n^\mathrm{L}$, we obtain the following simple expressions for integral charge and current at the edge:
\begin{align}
Q_{\mathrm{e}}&=-\frac{eg}{2\pi}\sum_n\Theta(E_\mathrm{F}^\mathrm{e}-E_n^\mathrm{L})k_n^\mathrm{max},\label{Q_e}\\
I_y&=-\frac{eg}{2\pi\hbar}\sum_n\Theta(E_\mathrm{F}^\mathrm{e}-E_n^\mathrm{L})(E_\mathrm{F}^\mathrm{e}-E_n^\mathrm{L}).\label{I_y}
\end{align}

\section{Theory of magnetoplasmons with edge charge and current}\label{Sec_WH}

Now we consider the problem of edge magnetoplasmons propagating along the straight boundary of 2D conducting material. Suppose the electric potential $\varphi$ and the oscillating charge density $\rho$ on the surface of the 2D material have the common plane-wave dependence on $y$ and $t$: $\varphi,\rho\propto e^{i(qy-\omega t)}$ (we assume  $q,\omega>0$). At each point $(x,y)$ of this material, the 2D current density $\mathbf{j}=\{j_x,j_y\}$ can be found as the matrix product of the local conductivity tensor $\sigma_{\alpha\beta}(x)$ and the electric field vector $\mathbf{E}=-\nabla\varphi=\{-\varphi'(x),-iq\varphi(x)\}$:
\begin{align}
j_x&=-\sigma_{xx}(x)\varphi'(x)-iq\sigma_{xy}(x)\varphi(x),\label{j_x}\\
j_y&=-\sigma_{yx}(x)\varphi'(x)-iq\sigma_{yy}(x)\varphi(x)+\Delta j_y^\mathrm{dr}(x).\label{j_y}
\end{align}
The key point here is the density of \emph{drift current} $\Delta j_y^\mathrm{dr}$ in the right hand side of Eq.~(\ref{j_y}), which is the oscillating part $\Delta j_y^\mathrm{dr}\propto e^{i(qy-\omega t)}$ of the boundary current (\ref{I_y}) carried by the edge channels and distributed in space near the sample boundary as shown in Fig.~\ref{Fig_dens_curr}(b). This term is absent in the conventional theory of edge modes \cite{Fetter,Wang,Aleiner,Johnson, Volkov,Margetis,Sokolik,Sokolik2}. The oscillations of drift current density $\Delta j_y^\mathrm{dr}$ originate from transient changes of the edge state occupations due to oscillations $\Delta E_\mathrm{F}^\mathrm{e}\propto e^{i(qy-\omega t)}$ of the local Fermi level at the edge $E_\mathrm{F}^\mathrm{e}=E_\mathrm{F}+\Delta E_\mathrm{F}^\mathrm{e}$ around its equilibrium bulk value $E_\mathrm{F}$, see Fig.~\ref{Fig_picture}. If the amplitude of these oscillations is relatively small, we assume the linear response of magnetic edge channels taking small oscillating perturbations of Eqs.~(\ref{Q_e})--(\ref{I_y}): $\Delta Q_\mathrm{e}=(\partial Q_\mathrm{e}/\partial E_\mathrm{F}^\mathrm{e})\Delta E_\mathrm{F}^\mathrm{e}$, $\Delta I_y=(\partial I_y/\partial E_\mathrm{F}^\mathrm{e})\Delta E_\mathrm{F}^\mathrm{e}$.

The oscillating integral edge current $\Delta I_y=\int_0^\infty dx\,\Delta j_y^\mathrm{dr}(x)$ is distributed over several magnetic lengths near the edge $x=0$, as seen from Fig.~\ref{Fig_dens_curr}(b). We can introduce the model distribution function $a(x)$, which is essentially nonzero near $x=0$ on the scale $x\sim l_H$ and normalized to unity, such as $j_y^\mathrm{dr}(x)=a(x)\Delta I_y$, $\int_{-\infty}^\infty dx\,a(x)=1$. The coordinate-dependent components $\sigma_{\alpha\beta}(x)$ of the conductivity tensor rise from zero in empty space $x<0$ to their bulk values $\sigma_{\alpha\beta}$ at $x\rightarrow\infty$. This also happens on the scale of several magnetic lengths, so, for simplicity of the following calculations, we can assume that $\partial_x\sigma_{\alpha\beta}(x)=a(x)\sigma_{\alpha\beta}$. As will be clear below, the exact shape of the function $a(x)$ is not important whilst the charge density $\rho(x)$ and electric potential $\varphi(x)$ profiles of magnetoplasmon oscillations are slowly varying on the scale of $l_H$. Thus, using the continuity equation $\partial\rho/\partial t+\mathrm{div}\,\mathbf{j}=0$, which in our case turns into $-i\omega\rho+\partial_x j_x+iqj_y=0$, we obtain from Eqs.~(\ref{j_x})--(\ref{j_y}):
\begin{align}
i\omega\rho(x)=a(x)\{-\sigma_{xx}\varphi'(x)-iq\sigma_{xy}\varphi(x)+iq\Delta I_y\}\nonumber\\
-iq(\sigma_{xy}+\sigma_{yx})\varphi'(x)-\sigma_{xx}\varphi''(x)+\sigma_{yy}q^2\varphi(x).\label{hydr1}
\end{align}
For isotropic material in magnetic field, $\sigma_{xx}=\sigma_{yy}$, $\sigma_{xy}=-\sigma_{yx}$, so the second term in the right hand side vanishes.

The oscillations $\Delta E_\mathrm{F}^\mathrm{e}$ of the local Fermi level at the edge can be related to those of the integral charge $\Delta Q_\mathrm{e}$, so that $\Delta I_y=(\partial I_y/\partial E_\mathrm{F}^\mathrm{e})(\partial Q_\mathrm{e}/\partial E_\mathrm{F}^\mathrm{e})^{-1}\Delta Q_\mathrm{e}$. In strong enough magnetic field, where $l_H$ is much smaller than characteristic length scale of $\varphi(x)$, which is known to be of the order of $q^{-1}$ \cite{Volkov,Sokolik2}, we can replace $a(x)$ by the Dirac delta function, $a(x)\approx\delta(x)$. Thus, Eq.~(\ref{hydr1}) takes the form
\begin{align}
i\omega\rho(x)=\delta(x)\left\{-\sigma_{xx}\varphi'(x)-iq\left[\sigma_{xy}-v_\mathrm{dr}\frac{\Delta Q_\mathrm{e}}{\varphi(0)}\right]\varphi(0)\right\}\nonumber\\
-\sigma_{xx}(\partial_x^2-q^2)\varphi(x),\label{hydr2}
\end{align}
where the drift velocity
\begin{equation}
v_\mathrm{dr}=\frac{\partial I_y}{\partial Q_\mathrm{e}}=\left.\frac{\partial I_y/\partial E_\mathrm{F}^\mathrm{e}}{\partial Q_\mathrm{e}/\partial E_\mathrm{F}^\mathrm{e}}\right|_{E_\mathrm{F}^\mathrm{e}=E_\mathrm{F}}\label{v_dr}
\end{equation}
was introduced. This velocity jointly characterizes the charge and density responses of the edge channels and can be calculated explicitly from Eqs.~(\ref{Q_e})--(\ref{I_y}). Assuming that the Fermi level lies between $n_0$th and $(n_0+1)$th bulk Landau levels, $n_0+\frac12<E_\mathrm{F}/\hbar\omega_\mathrm{c}<n_0+\frac32$, so that $n_0$ edge channels contribute to the sums in Eqs.~(\ref{Q_e})--(\ref{I_y}), we obtain
\begin{equation}
v_\mathrm{dr}=\frac{n_0/\hbar}{\displaystyle\sum_{n=0}^{n_0}\frac{\partial k_n^\mathrm{max}}{\partial E_{nk_y}}}=\left\langle\left(\frac1\hbar\frac{\partial E_{nk_y}}{\partial k_y}\right)^{-1}\right\rangle^{-1},
\label{v_dr_harm}
\end{equation}
where the average $\langle\ldots\rangle$ is taken at the Fermi level $E_{nk_y}=E_\mathrm{F}$ over all edge states crossing this level. Thus we may interpret $v_\mathrm{dr}$ as a \emph{harmonic mean of group velocities} of all magnetic edge channels crossing the Fermi level.

Eq.~(\ref{hydr2}) has a form of conventional continuity equation in the theory of plasmon-type edge modes \cite{Wang,Johnson,Aleiner,Volkov,Sokolik,Sokolik2}, but with the Hall conductivity $\sigma_{xy}$ replaced by
\begin{equation}
\tilde\sigma_{xy}=\sigma_{xy}-v_\mathrm{dr}\frac{\Delta Q_\mathrm{e}}{\varphi(0)}\label{sigma_ren}
\end{equation}
due to the response of edge channels. Since in magnetic field along the positive $z$ axis with negatively charged carriers $\sigma_{xy}<0$ at low frequencies $\omega\rightarrow0$, and $\Delta Q_\mathrm{e}/\varphi(0)>0$, the edge channels effectively increase the negative $\sigma_{xy}$ by the absolute value. In other words, the current response of edge channels acts in phase with the bulk currents caused by the Hall conductivity, as depicted in Fig.~\ref{Fig_picture}. As will be shown below [see Eq.~(\ref{disp_approx})], in the long-wavelength limit the magnetoplasmon phase velocity $\omega/q$ is approximately proportional to $|\sigma_{xy}|$ \cite{Volkov,Sokolik,Margetis}, thus the edge channel response speeds up the magnetoplasmons.

We need to supplement Eq.~(\ref{hydr2}) with the Poisson equation $\varepsilon_\mathrm{b}\nabla^2\varphi=-4\pi\rho\delta(z)$, where $\varepsilon_\mathrm{b}$ is the background dielectric constant of the surrounding medium. For a half-plane material, this equation at $z=0$ can be transformed into the integral form \cite{Fetter,Volkov,Sokolik,Margetis}
\begin{equation}
\varphi(x)=\frac{4\pi}{\varepsilon_\mathrm{b}}\int\limits_0^\infty dx'\:L(x-x')\rho(x'),\label{Poisson}
\end{equation}
where the kernel
\begin{equation}
L(x-x')=\int\limits_{-\infty}^\infty\frac{dk}{2\pi}\frac{e^{ik(x-x')}}{2\sqrt{k^2+q^2}}=\frac1{2\pi}K_0(q|x-x'|)\label{kernel}
\end{equation}
is given in terms of the second kind Bessel function $K_0$.

The system of equations (\ref{hydr2}), (\ref{Poisson}) can be solved analytically using the Wiener-Hopf method \cite{Volkov,Sokolik,Sokolik2,Margetis,Margetis2,Carrier} after the Fourier transforms of $\varphi(x)$ and $\rho(x)$ carried out separately at $x<0$ and $x>0$: $\Phi_+(\xi)=\int_0^\infty dx\,e^{iq\xi x}\varphi(x)$, $\Phi_-(\xi)=\int_{-\infty}^0dx\,e^{iq\xi x}\varphi(x)$, $Q_+(\xi)=\int_0^\infty dx\,e^{iq\xi x}\rho(x)$. Applying these transforms to Eqs.~(\ref{hydr2}), (\ref{Poisson}), we obtain
\begin{align}
i\omega &Q_+(\xi)=-iq\varphi(0)(\xi\sigma_{xx}+\tilde\sigma_{xy})+q^2\sigma_{xx}(\xi^2+1)\Phi_+(\xi),\label{WH_eq1}\\
&\Phi_+(\xi)+\Phi_-(\xi)=\frac{4\pi}{\varepsilon_\mathrm{b}q}L(\xi)Q_+(\xi),\label{WH_eq2}
\end{align}
where, according to Eq.~(\ref{kernel}), the dimensionless Fourier-transformed kernel is $L(\xi)=[2\sqrt{\xi^2+1}]^{-1}$. Excluding $Q_+$ from Eqs.~(\ref{WH_eq1})--(\ref{WH_eq2}), we obtain
\begin{equation}
\varepsilon(\xi)\Phi_+(\xi)+\Phi_-(\xi)=-\frac{i\varphi(0)}qL(\xi)(\eta\xi-i\tilde\chi),\label{WH_eq3}
\end{equation}
where the dielectric function of the 2D material
\begin{equation}
\varepsilon(\xi)=1-\eta L(\xi)(\xi^2+1)=1-\frac\eta2\sqrt{\xi^2+1},\label{epsilon}
\end{equation}
at the wave vector $\mathbf{k}=\{-q\xi,q\}$ is introduced. The dimensionless conductivities
\begin{equation}
\eta=\frac{4\pi q\sigma_{xx}}{i\varepsilon_\mathrm{b}\omega},\quad
\chi=\frac{4\pi q\sigma_{xy}}{\varepsilon_\mathrm{b}\omega}\label{eta_chi}
\end{equation}
are purely real in the static nondissipative limit, when $\sigma_{xx}$ and $\sigma_{xy}$ are, respectively, imaginary and real. Since $\sigma_{xy}$ is renormalized, see Eq.~(\ref{sigma_ren}), due to the edge channel response, $\chi$ gets renormalized as well:
\begin{equation}
\tilde\chi=\chi-\frac{4\pi q v_\mathrm{dr}}{\varepsilon_\mathrm{b}\omega}\frac{\Delta Q_\mathrm{e}}{\varphi(0)}.\label{chi_ren}
\end{equation}

According to definitions, the functions $\Phi_+(\xi)$ and $\Phi_-(\xi)$ are analytical in, respectively, upper and lower half-planes of the complex dimensionless wave vector $\xi$. The Wiener-Hopf method \cite{Volkov,Carrier} relies on the decomposition $\varepsilon(\xi)=F_+(\xi)/F_-(\xi)$ of the dielectric function (\ref{epsilon}), where $F_\pm(\xi)$ are analytical in, respectively, upper and lower half-planes. This decomposition can be carried out analytically (see also Ref. \cite{Nikulin}), and the result is
\begin{align}
F_+(\xi)&=-\frac\eta2\frac{(1-z_1)^{(s+1)/2}(1+z_2)^{1/2}}{(2p)^{1/2}(1+z_1)^{s/2}}K(\xi),\label{Fp_an}\\
F_-(\xi)&=\frac{(2p)^{1/2}(1+z_2)^{s/2}}{(1+z_1)^{1/2}(1-z_2)^{(s+1)/2}}K(\xi),\label{Fm_an}\\
K(\xi)&=\exp\left\{\frac{\pi i}4+\frac{s}{2\pi i}\left[f(z_1)+f(z_2)\right]\right\},\\
f(z)&=-\frac{\pi^2}6+\log z\log(1+z)\nonumber\\
&\hphantom{=}\:\,+\mathrm{Li}_2(-z)+\mathrm{Li}_2(1-z),
\end{align}
where $z_1=isp\lambda$, $z_2=-isp/\lambda$, $s=\mathrm{sign}(\mathrm{Re}\,\eta)$, $p=\xi+\sqrt{\xi^2+1}$, $\lambda=-2i/\eta+\sqrt{1-(2/\eta)^2}$; $\mathrm{Li}_2$ is the dilogarithm function. These formulas are applicable for any complex $\eta$, i.e. in both capacitive ($\mathrm{Im}\,\sigma_{xx}<0$, $\mathrm{Re}\,\eta<0$) and inductive ($\mathrm{Im}\,\sigma_{xx}>0$, $\mathrm{Re}\,\eta>0$) regimes of the conductivity, and for both materials with dissipation ($\mathrm{Re}\,\sigma_{xx}>0$, $\mathrm{Im}\,\eta<0$) and active media ($\mathrm{Re}\,\sigma_{xx}<0$, $\mathrm{Im}\,\eta>0$).

Equating separately the parts of Eq.~(\ref{WH_eq3}), which are analytical in upper and lower complex half-planes of $\xi$, we obtain the Fourier transforms of the potential (see the details in \cite{Volkov,Sokolik2}):
\begin{align}
\Phi_\pm(\xi)=\pm\frac{i\varphi(0)}{2q}&\left\{\frac{1-\tilde\chi/\eta}{\xi-i}\left[1-\frac{F_+(i)}{F_\pm(\xi)}\right]\right.\nonumber\\
&\left.+\frac{1+\tilde\chi/\eta}{\xi+i}\left[1-\frac{F_-(-i)}{F_\pm(\xi)}\right]\right\}.\label{WH_Phi}
\end{align}
At $|\xi|\rightarrow\infty$ the functions (\ref{Fp_an})--(\ref{Fm_an}) behave as $F_\pm(\xi)\propto\xi^{\pm1/2}$, which allows us to find the power-law asymptotics of $\Phi_\pm(\xi)$ at large $\xi$. They are related to the limiting behavior of the potential $\varphi(x)=q\int_{-\infty}^\infty d\xi\,e^{-iq\xi x}[\Phi_+(x)+\Phi_-(x)]$ at $x\rightarrow0$. By equating the limiting value $\varphi(x\rightarrow0)$ to $\varphi(0)$ in Eq.~(\ref{WH_Phi}), we find the dispersion equation for edge magnetoplasmons:
\begin{equation}
(\eta-\tilde\chi)F_+(i)+(\eta+\tilde\chi)F_-(-i)=0.\label{WH_Disp1}
\end{equation}
Using the analytical expressions (\ref{Fp_an})--(\ref{Fm_an}), this equation can be written explicitly as:
\begin{equation}
\frac{F_+(i)}{F_-(-i)}=\frac{1+\lambda}{1-\lambda}\exp \left\{\frac{2i}\pi f(\lambda)\right\}=\frac{\tilde\chi+\eta}{\tilde\chi-\eta}.\label{WH_Disp2}
\end{equation}

In order to calculate the magnetoplasmon dispersion from Eq.~(\ref{WH_Disp2}) with taking into account the edge channel response (\ref{chi_ren}), we need to find the ratio $\Delta Q_\mathrm{e}/\varphi(0)$, which defines the fraction of a total magnetoplasmon oscillating charge which is accommodated by the edge channels. The theory of edge modes \cite{Volkov,Sokolik2} shows that oscillating charge density $\rho(x)$ behaves as $\rho(x)\propto x^{-1/2}$ at the distances from the edge $x=0$ much smaller than the wavelength $2\pi/q$. Thus we can assume that the dominating part of the charge is concentrated in very narrow strip near the edge, so $\Delta Q_\mathrm{e}\approx\int_0^\infty dx\,\rho(x)=Q_+(0)$. From Eqs.~(\ref{WH_eq2}), (\ref{WH_Phi}), and (\ref{WH_Disp1}) we can find the Fourier transform of charge density
\begin{equation}
Q_+(\xi)=\frac{\varphi(0)\varepsilon_\mathrm{b}}{4\pi}(\eta-\tilde\chi)\frac{F_+(i)}{F_+(\xi)}.\label{WH_Qp}
\end{equation}
Using the property $F_+(\xi)F_-(-\xi)=-\eta/2$ of the functions $F_\pm(\xi)$ and the dispersion equation (\ref{WH_Disp1}), we obtain  $F_+(0)=-(\eta/2)\sqrt{1-2/\eta}$ and $F_+(i)=\sqrt{\eta(\eta+\tilde\chi)/2(\eta-\tilde\chi)}$ (we take into account here that $\eta,\tilde\chi<0$ for a system in the quantum Hall regime at low frequencies of edge magnetoplasmons). Thus Eq.~(\ref{WH_Qp}) at $\xi=0$ results in
\begin{equation}
\Delta Q_\mathrm{e}=Q_+(0)=\frac{\varphi(0)\varepsilon_\mathrm{b}}{4\pi}\frac{\sqrt{\tilde\chi^2-\eta^2}}{\sqrt{1-\eta/2}}.\label{WH_Q0}
\end{equation}
Substituting Eq.~(\ref{WH_Q0}) to Eq.~(\ref{chi_ren}), we obtain
\begin{equation}
\tilde\chi=\chi-\frac{qv_\mathrm{dr}}\omega\frac{\sqrt{\tilde\chi^2-\eta^2}}{\sqrt{1-\eta/2}}.\label{chi_ren2}
\end{equation}
The equations (\ref{WH_Disp2}), (\ref{chi_ren2}), taken with the notations (\ref{v_dr}), (\ref{eta_chi}), and with the edge channel charge and current response functions (\ref{Q_e})--(\ref{I_y}), are the main result of this paper. These equations determine the edge magnetoplasmon dispersion $\omega(q)$, when the edge channel response, characterized by the drift velocity $v_\mathrm{dr}$, is taken into account.

\section{Edge magnetoplasmon dispersions}\label{Sec_calc}

From the point of view of the existing experiments on edge magnetoplasmons \cite{Talyanskii1,Ernst,Mast,Talyanskii2,Kumada3,Kumada4}, only the long-wavelength asymptotic of the dispersion relation $\omega(q)$ at $q,\omega\rightarrow0$ is usually relevant and observable. In this limit, for a conducting material with negligible dissipation, which turns into an insulator in the quantum Hall regime, the parameter $\eta=4\pi q\sigma_{xx}/i\varepsilon_\mathrm{b}\omega$ is small by the absolute value and negative. We can represent it as $\eta=-2qw$, where $w=-(2\pi/\varepsilon_b)(\partial\mathrm{Im}\sigma_{xx}/\partial\omega)|_{\omega=0}$ is the characteristic penetration length of the charge density oscillations (the 2D counterpart of a 3D conductor skin depth) \cite{Volkov,Sokolik}. The parameter $\chi=4\pi q\sigma_{xy}/\varepsilon_\mathrm{b}\omega$ remains finite and negative at $q,\omega\rightarrow0$. The asymptotic expressions for the functions (\ref{Fp_an})--(\ref{Fm_an}) at $\eta\rightarrow0$ can be written as:
\begin{align}
F_\pm(\xi)=\sqrt{-\frac\eta2}\left\{1-\frac{i\eta}{4\pi p}\left[(p^2+1)\left(\mp\frac{\pi i}2+\log p\right)\right.\right.\nonumber\\
\left.\left.+(p^2-1)\log\left(-\frac\eta{4\mathfrak{e}}\right)\vphantom{\frac{\pi i}2}\right]+\mathcal{O}(\eta^2)\right\},\label{F_asympt}
\end{align}
where $\mathfrak{e}$ is the base of a natural logarithm. Using these asymptotics, we obtain
\begin{equation}
\frac{F_+(i)}{F_-(-i)}\approx1+\frac\eta\pi\log\left(-\frac\eta{4\mathfrak{e}}\right),
\end{equation}
so we bring the dispersion equation (\ref{WH_Disp2}) to the approximate form
\begin{equation}
\frac1{\tilde\chi-\eta}=\frac1{2\pi}\log\left(-\frac\eta{4\mathfrak{e}}\right).\label{WH_Disp3}
\end{equation}
Taking into account that $|\eta|\ll|\tilde\chi|$ at $\omega\rightarrow0$ and using Eqs.~(\ref{eta_chi}), (\ref{chi_ren2}), we obtain the magnetoplasmon dispersion in the long-wavelength limit:
\begin{equation}
\omega\approx v_\mathrm{dr}q-\frac{2\sigma_{xy}q}{\varepsilon_\mathrm{b}}\log\frac{2\mathfrak{e}}{qw}.\label{disp_approx}
\end{equation}
This formula differs by the drift term $v_\mathrm{dr}q$ from the well-known asymptotic formula for edge magnetoplasmon dispersion \cite{Volkov,Sokolik}. Thus, due to additional current response of the magnetic edge channels, the edge magnetoplasmon phase velocity $v=\omega/q$ increases by $v_\mathrm{dr}$.

Although our analysis started from the hard-edge boundary conditions, this result is surprisingly consistent with the classical picture of the soft edge formed by a local electric field $\mathcal{E}_x^\mathrm{conf}$ which confines electrons inside the sample. Indeed, such field gives rise to a spiraling motion of the electrons in magnetic field with the \emph{classical} drift velocity $v_\mathrm{dr}^\mathrm{c}=-c\mathcal{E}_x^\mathrm{conf}/B$ along the edge. From the other hand, the edge state wave function is typically concentrated near $x_0=-l_H^2k_y$ (at least, when $k_y\rightarrow-\infty$, see Fig.~\ref{Fig_wf}), so we can relate the $x_0$- and $k_y$-derivatives of edge state energies to obtain the effective confining field in terms of mean force $F_x=-e\mathcal{E}_x^\mathrm{conf}=-\partial E_{nk_y}/\partial x_0\approx l_H^{-2}\partial E_{nk_y}/\partial k_y$ acting on electrons at the Fermi level near the edge. Hence the classical drift velocity is $v_\mathrm{dr}^\mathrm{c}=\hbar^{-1}(\partial E_{nk_y}/\partial k_y)$, which agrees with the quantum-mechanical expression (\ref{v_dr_harm}) if we neglect the variance of edge state group velocities.

The \emph{quasiclassical} approximation for $v_\mathrm{dr}$ can be obtained from Eq.~(\ref{v_dr_harm}) in the limit when a large number of bulk Landau levels $n_0=E_\mathrm{F}/\hbar\omega_\mathrm{c}\gg1$ is occupied, and when we find the edge state energies $E_{nk_y}$ approximately using the quasiclassical quantization condition. As shown in Appendix~\ref{Appendix_A},
\begin{equation}
v_\mathrm{dr}^\mathrm{qc}=l_H\omega_\mathrm{c}\sqrt{\frac{n_0}2}=\sqrt{\frac{E_\mathrm{F}}{2m}}.\label{v_dr_qc}
\end{equation}
Assuming that the Fermi energy $E_\mathrm{F}=\frac12mv_\mathrm{F}^2$ can be related to the Fermi velocity $v_\mathrm{F}$ of electrons at the edge, we obtain $v_\mathrm{dr}^\mathrm{qc}=\frac12v_\mathrm{F}$. This estimate agrees with Ref.~\cite{Montambaux} where the velocity of quasiclassical motion on skipping orbits in magnetic field was found to be of the order of the Fermi velocity, and with the recent experiment \cite{Martinez}.

\begin{figure}[t]
\begin{center}
\includegraphics[width=\columnwidth]{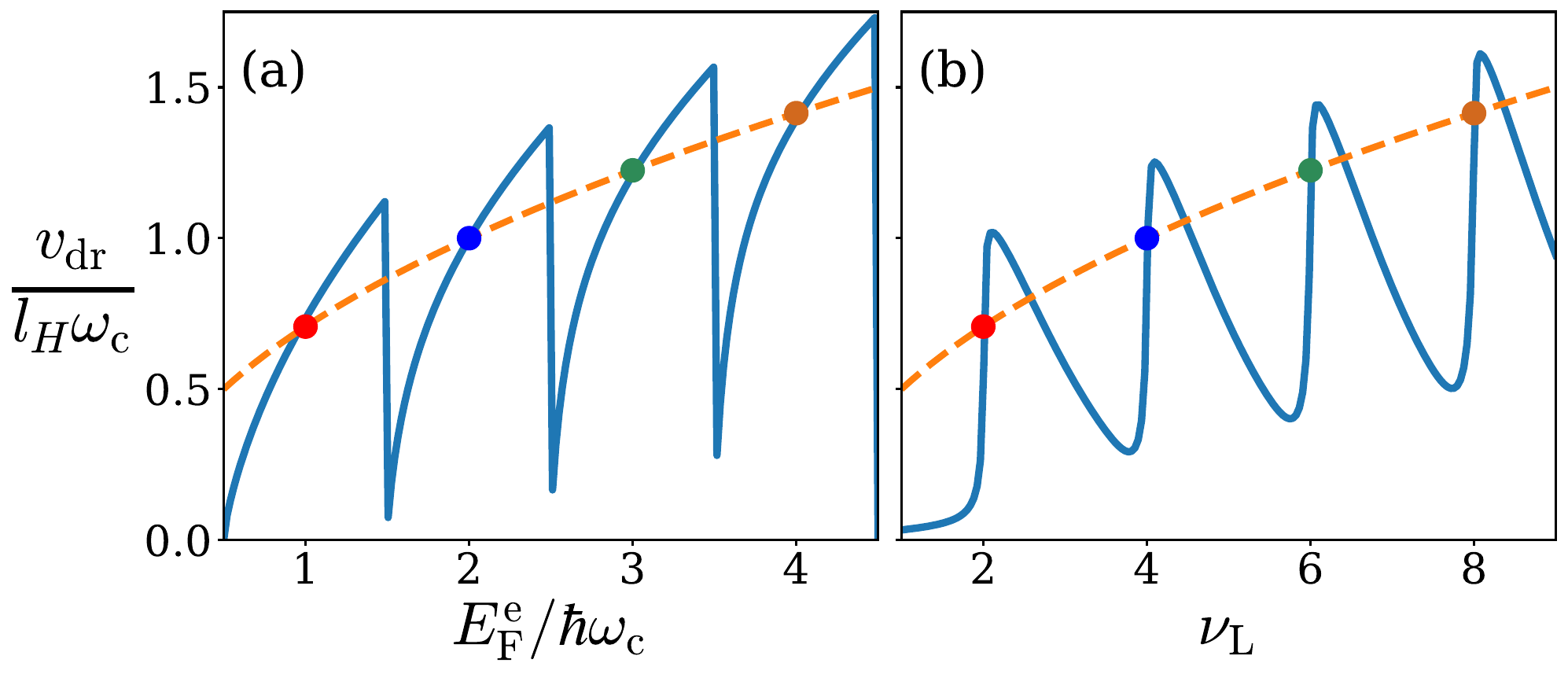}
\end{center}
\caption{\label{Fig_v_dr}Drift velocity $v_\mathrm{dr}$ (solid lines) as function of (a) Fermi level $E_\mathrm{F}$ and (b) Landau level filling factor $\nu_\mathrm{L}$. Dashed lines show the quasiclassical approximation (\ref{v_dr_qc}) which coincide (circles) with the exact quantum-mechanical $v_\mathrm{dr}$ when the Fermi level is located in the mid-gap between Landau levels, i.e. when $E_\mathrm{F}/\hbar\omega_\mathrm{c}$ and $\frac12\nu_\mathrm{L}$ are equal integers.}
\end{figure}

The drift velocity $v_\mathrm{dr}$, calculated according to quantum-mechanical formulas (\ref{v_dr}) and (\ref{Q_e})--(\ref{I_y}), is shown in Fig.~\ref{Fig_v_dr} as function of the Fermi level $E_\mathrm{F}$ and Landau level filling factor $\nu_\mathrm{L}$, i.e. number of occupied Landau levels with taking into account their double spin degeneracy $g$. To calculate the $\nu_\mathrm{L}$-dependence, a weak broadening of the Landau levels is assumed, which makes the dependence of $E_\mathrm{F}$ on the electron density $n_\mathrm{e}=\nu_\mathrm{L}/2\pi l_H^2$ almost step-like, but smooth, when the system goes through a sequence of interchanging compressible and incompressible phases \cite{Talyanskii1,Chklovskii}. As seen in Fig.~\ref{Fig_v_dr}, the quasiclassical approximations (\ref{v_dr_qc}) coincides with the exact result (\ref{v_dr}) when the Landau level filling is strictly integer, i.e. in the incompressible phases when the Fermi level is located in the mid-gaps. In the following, we will consider only such cases of integer Landau level fillings corresponding to the quantum Hall regime, when $E_\mathrm{F}/\hbar\omega_\mathrm{c}=n_0+1=\nu_\mathrm{L}/g$.

To calculate the conductivity tensor $\sigma_{\alpha\beta}(\omega)$ of 2DEG in magnetic field, we can use the approach similar to those applied to graphene in Ref.~\cite{Sokolik}. In the $\mathbf{E}\cdot\mathbf{r}$ gauge, with the assumption of Lorentzian broadening of the spectral function $A_n(E)=(\Gamma/\pi)/[(E-E_n^\mathrm{L})^2+\Gamma^2]$ of each $n$th Landau level with the width $\Gamma$, we obtain
\begin{align}
\sigma_{xx}(\omega)&=\frac{ige^2\hbar\omega_\mathrm{c}^2}{4\pi}\sum_{n=1}^\infty n[I_{n,n-1}(\omega)+I_{n-1,n}(\omega)],\\
\sigma_{xy}(\omega)&=\frac{ge^2\hbar\omega_\mathrm{c}^2}{4\pi}\sum_{n=1}^\infty n[I_{n,n-1}(\omega)-I_{n-1,n}(\omega)],
\end{align}
where
\begin{align}
I_{n_1n_2}(\omega)&=\int dE_1dE_2\:A_{n_1}(E_1)A_{n_2}(E_2)\nonumber\\
&\times\frac{f(E_2)-f(E_1)}{(E_1-E_2)(\hbar\omega+E_2-E_1+i\delta)},\label{I_n1n2}
\end{align}
and $f(E)=\Theta(E_\mathrm{F}-E)$ is the low-temperature occupation number of electronic state with the energy $E$. The integrals (\ref{I_n1n2}) take into account both inter-Landau-level electron transitions and intralevel transitions inside each broadened Landau level, in case of its partial filling. In the case of integer Landau level filling $E_\mathrm{F}=\hbar\omega_\mathrm{c}(n_0+1)$, and when the energy width of each level is much smaller than the cyclotron energy, $\Gamma\ll\hbar\omega_\mathrm{c}$ (clean limit), we obtain $I_{n,n-1}(\omega)=\delta_{n,n_0+1}/\hbar^2\omega_\mathrm{c}(\omega-\omega_\mathrm{c})$, $I_{n-1,n}(\omega)=\delta_{n,n_0+1}/\hbar^2\omega_\mathrm{c}(\omega+\omega_\mathrm{c})$.The resulting conductivities of 2DEG in magnetic field calculated in the clean limit in the insulating (or incompressible) regime coincide with the classical Drude result:
\begin{equation}
\sigma_{xx}(\omega)=\frac{i\nu_\mathrm{L} e^2}{2\pi\hbar}\frac{\omega\omega_\mathrm{c}}{\omega^2-\omega_\mathrm{c}^2},\quad
\sigma_{xy}(\omega)=\frac{\nu_\mathrm{L} e^2}{2\pi\hbar}\frac{\omega_\mathrm{c}^2}{\omega^2-\omega_\mathrm{c}^2}.
\end{equation}
The low-frequency asymptotics of these expressions $\sigma_{xx}\approx-(\varepsilon_\mathrm{b}/2\pi)i\omega w$, $w=\nu_\mathrm{L} e^2/\varepsilon_\mathrm{b}\hbar\omega_\mathrm{c}$, $\sigma_{xy}\approx-\nu_\mathrm{L} e^2/2\pi\hbar$ can be used in Eq.~(\ref{disp_approx}) to find an approximate magnetoplasmon dispersion in the long-wavelength limit.

Calculation results for the edge magnetoplasmon dispersion in our approach depend on the dimensionless parameter
\begin{equation}
r_\mathrm{s}=\frac{e^2/\varepsilon_\mathrm{b}l_H}{\hbar\omega_\mathrm{c}},
\end{equation}
which defines the ratio of characteristic Coulomb interaction and kinetic energies of electrons in magnetic field. As follows from Eqs.~(\ref{disp_approx}) and (\ref{v_dr_qc}), the ratio of $v_\mathrm{dr}$ to the typical magnetoplasmon velocity $v=\omega/q\sim-\sigma_{xy}/\varepsilon_\mathrm{b}$ found in the absence of the drift contribution, behaves as $v_\mathrm{dr}/v\sim1/r_\mathrm{s}$. Since $r_\mathrm{s}\propto1/\varepsilon_\mathrm{b}\sqrt{B}$, we expect significant contribution of $v_\mathrm{dr}$ to the total velocity of edge magnetoplasmon at strong dielectric screening or in strong magnetic field. In experiments on edge magnetoplasmons dealing with GaAs-based quantum wells \cite{Talyanskii1,Ernst,Mast,Talyanskii2}, the system parameters are typically $\varepsilon_\mathrm{b}=4-13$, $B=4-20\,\mbox{T}$, and the electron effective mass is $m=0.067m_0$, so $r_\mathrm{s}\approx0.6-4$. In the case of graphene \cite{Kumada3,Kumada4}, the role of cyclotron energy $\hbar\omega_\mathrm{c}$ is played by the characteristic distance between Landau level $\hbar v_\mathrm{F}/l_H$, where $v_\mathrm{F}\approx10^6\,\mbox{m/s}$, so $r_\mathrm{s}$ does not depend on $B$ but, depending on the substrate dielectric constant $\varepsilon_\mathrm{b}$, can take the values $r_\mathrm{s}\sim2$ (for suspended graphene) or less.

\begin{figure}[t]
\begin{center}
\includegraphics[width=\columnwidth]{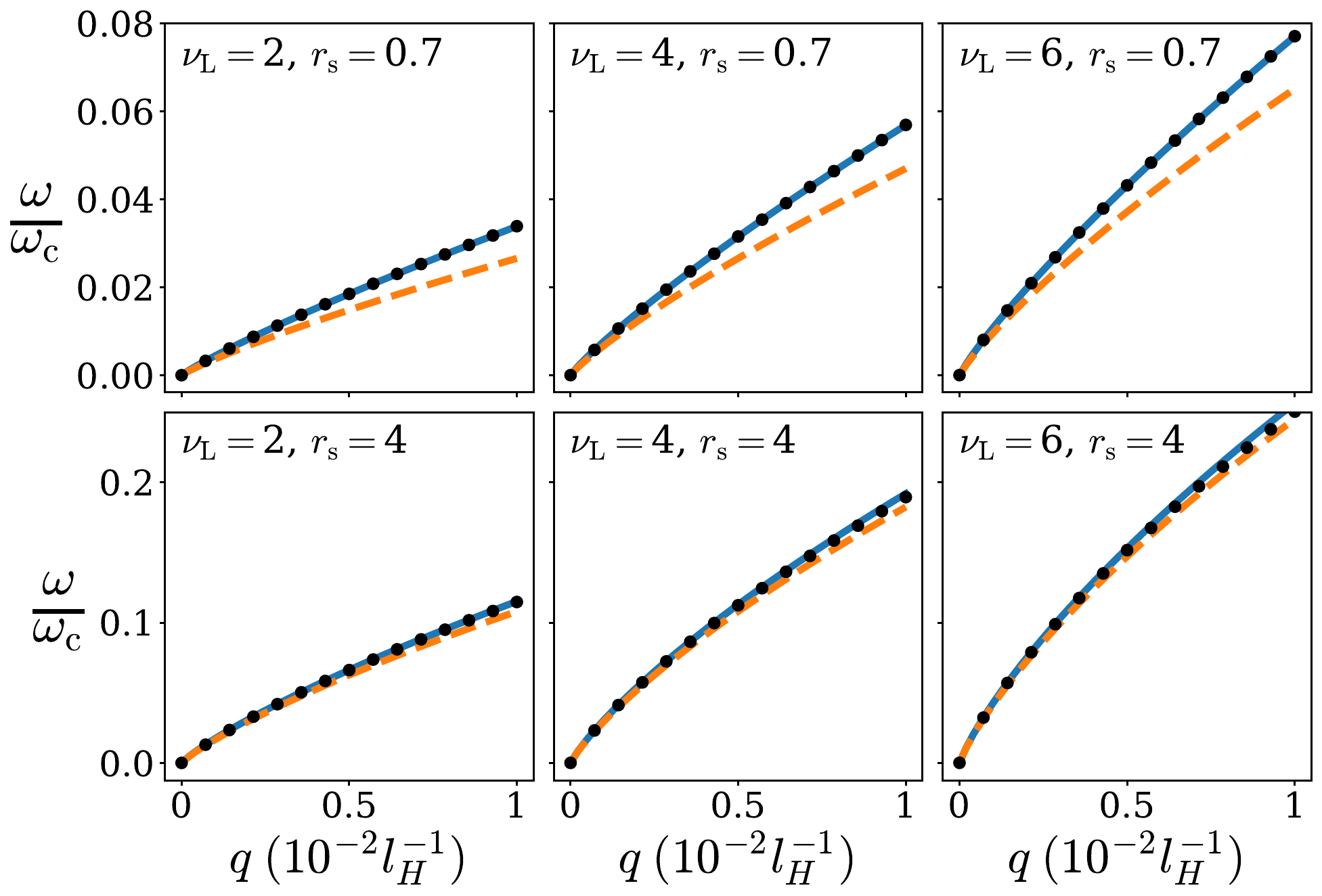}
\end{center}
\caption{\label{Fig_disp}Edge magnetoplasmon dispersions found numerically in the presence (solid lines) and absence (dashed lines) of the drift velocity $v_\mathrm{dr}$. Black circles show the analytical approximation (\ref{disp_approx}) for the long-wavelength limit. The panels correspond to different Landau level filling factors $\nu_\mathrm{L}$ and Coulomb interaction strengths $r_\mathrm{s}$.}
\end{figure}

Fig.~\ref{Fig_disp} shows the example results (solid lines) of numerical solution of the dispersion equation (\ref{WH_Disp2}) with taking into account the contribution of the drift velocity. In experiments \cite{Wassermeier,Ashoori,Talyanskii1,Ernst,Mast,Talyanskii2,Kamata,Kumada1,Andreev,Wu,Endo,Kumada2,Crassee,Yan,Kumada3,Kumada4,Petkovic1,Lin,Petkovic2,Kumada5}, typical wavenumbers $q$ of edge magnetoplasmons are $10-10^4\,\mbox{cm}^{-1}$, while magnetic length is $l_H\sim10\,\mbox{nm}$ at $B=4-20\,\mbox{T}$, so the dimensionless product $ql_H$ does not exceed $10^{-2}$. As seen in Fig.~\ref{Fig_disp}, in this regime the long-wavelength asymptotic (\ref{disp_approx}), shown by black circles, proves to be quite accurate. For the sake of comparison, the dispersions calculated in the absence of $v_\mathrm{dr}$ are shown by dashed lines. The relative role of $v_\mathrm{dr}$ increases with decrease of $r_\mathrm{s}$, in agreement with the arguments presented above. The total velocity of magnetoplasmons rises with increase of the filling factor $\nu_\mathrm{L}$, which agrees with experimental data \cite{Ernst,Talyanskii2,Kumada1,Kumada2}.

We have considered clean system, however our analysis of the role of drift velocity remains valid in the presence of dissipation as well. Nonzero DC limit of conductivity $\lim_{\omega\rightarrow0}\sigma_{xx}(\omega)=\sigma_\mathrm{DC}$ implies that $\eta$ acquires finite value $\eta=-4\pi iq\sigma_\mathrm{DC}/\varepsilon_\mathrm{b}$ instead of being infinitesimally small. Since $\eta$ is still much smaller by absolute value than $\tilde\chi$ in the strong magnetic field limit, it enters only the logarithm in the dispersion equation (\ref{WH_Disp3}), so the approximation $\omega\approx-2\sigma_{xy}q/\varepsilon_\mathrm{b}$ [see Eq.~(\ref{disp_approx})] would be sufficient to estimate $\eta\approx2\pi i\sigma_\mathrm{DC}/\sigma_{xy}$. The resulting real part of long-wavelength dispersion
\begin{equation}
\omega\approx v_\mathrm{dr}q-\frac{2\sigma_{xy}q}{\varepsilon_\mathrm{b}}\log\frac{2\mathfrak{e}|\sigma_{xy}|}{\pi\sigma_\mathrm{DC}}\label{disp_approx_dirty}
\end{equation}
has the same form as in the clean limit (\ref{disp_approx}) if the characteristic screening length is assumed to be $w=2\pi\sigma_\mathrm{DC}/\varepsilon_\mathrm{b}\omega\approx\pi\sigma_\mathrm{DC}/q|\sigma_{xy}|$. The same expression for $w$ in the presence of significant dissipation was derived in Ref.~\cite{Volkov}. Thus the additive contribution of $v_\mathrm{dr}$ to the total edge magnetoplasmon velocity arises in the long-wavelength limit irrespective of the role of dissipation.

\section{Conclusions}\label{Sec_concl}

Starting from quantum-mechanical analysis of magnetic edge channels, we developed consistent theory describing the drift velocity of edge magnetoplasmons. The unidirectional propagation of electrons populating the edge channels causes additional current response emerging during magnetoplasmon oscillations and accompanying perturbations of the local Fermi level near the edge. This response gives rise to additional term in the equations which describe edge magnetoplasmon electrodynamics. The magnitude of this term is characterized by the drift velocity $v_\mathrm{dr}$ calculated analytically as harmonic mean of group velocities of all magnetic edge channels whose dispersions cross the Fermi level. At the quasiclassical level, $v_\mathrm{dr}$ is analogous to the average velocity of electron spiraling motion in a confining electric field near the edge and in perpendicular magnetic field. As we show in Appendix~\ref{Appendix_A} using the quasiclassical quantization, $v_\mathrm{dr}$ is close to one half of the electron Fermi velocity when large number of Landau levels is occupied.

Due to Landau quantization, the quantum-mechanically calculated $v_\mathrm{dr}$ strongly oscillates when the Landau level filling factor is changed, but on average it is close to one half of electron Fermi velocity, in agreement with the classical picture of skipping orbits. Assuming that charge density oscillations are confined in a narrow strip near the edge, we solved analytically the problem of edge magnetoplasmon in the presence of the drift velocity using the Wiener-Hopf method. We show that in the low-frequency and long-wavelength limit, $v_\mathrm{dr}$ provides additive contribution to the edge magnetoplasmon velocity, which agrees with traditional treatment \cite{Kumada3,Kumada4} based on transition to a reference frame moving with the drift velocity $v_\mathrm{dr}$. Our conclusion about additive contribution of the drift velocity holds both in the clean limit and in the presence of dissipation.

The presented analysis, where the Dirichlet boundary condition is imposed on electron wave functions, corresponds to a thin-film or 2D material with an abrupt edge, examples are etched semiconductor quantum wells, atomically thin transition metal dichalcogenides \cite{Wang_TMDC}, and other 2D conducting and semiconducting materials such as phosphorene \cite{Kezerashvili} or TiN \cite{Shah}. The drift velocity of edge magnetoplasmons in such physical realizations with the hard edge was not considered theoretically so far. Nevertheless, in the classical limit our results come to surprising agreement with the opposite picture of smooth edge created by confining electric field.

Out theoretical analysis opens the door for prediction of edge-mode velocities with taking into account the local edge response beyond long-wavelength phenomenological and classical approaches. Numerical calculations for specific physical realizations will be performed in the future. Further development of the theory adapted to more complicated electronic structure and electromagnetic response near the edge should take into account appearance of dissipation at non-integer Landau level fillings \cite{Johnson,Sokolik}, and spatial structure of compressible and incompressible stripes \cite{Talyanskii1,Chklovskii}. Anisotropy of electron dispersion arising due to intrinsic properties of a material (such as AlAs \cite{Shayegan}) or fabrication-related strain can also affect the edge state and edge magnetoplasmon physics. After rescaling of coordinates, anisotropy of effective mass becomes equivalent to anisotropic Coulomb interaction, which affects integer and fractional quantum Hall effects \cite{Balagurov,Shayegan,Ciftja,Jiang} uncovering the physics of electron nematics \cite{Fradkin}. Edge mode electrodynamics becomes more mathematically involved in anisotropic case \cite{Sokolik2}, so interplay of drift velocity and anisotropy deserves additional study. Multiple-subband effects can modify the structure of Landau levels triggering their crossing \cite{Zhang,Ellenberger}, and spin-orbit interaction arising at the interfaces where 2DEG is formed \cite{Khalsa,Zhou} can also modify the structure of Landau levels in the bulk \cite{Wang2003,Zhang2006} and near the edge \cite{Grigoryan}. These effects can change the edge magnetoplasmon properties and electromagnetic response of magnetic edge channels at specific Landau level filling factors, however we expect the qualitative features of drift velocity to be robust against band structure modifications at low energy scale, because our results demonstrate correct quasiclassical limiting behavior arising when large number of Landau levels is occupied.

It is also of interest to extend our theory on massless Dirac electrons in doped graphene with accounting for specific character of their edge states at boundaries of graphene lattice having different orientations \cite{Abanin}, and on different unconventional materials such as quantum anomalous Hall insulators \cite{Mahoney2017,Wang2023,Martinez}, magic-angle twisted bilayer graphene \cite{Do} with a flat-band electron spectrum, and other graphene-based heterostructures.

\section*{Acknowledgments}
The work was supported by the Russian Science Foundation (grant No. 23-12-00115). Part of the work aimed on calculating the edge channel electromagnetic response was supported by Foundation for the Advancement of Theoretical Physics and Mathematics``BASIS''.

\appendix

\section{Quasiclassical approximation\\ for drift velocity}\label{Appendix_A}

The problem (\ref{Schrodinger}) of one-dimensional motion in harmonic potential with hard wall
\begin{equation}
V(x)=\left\{
\begin{array}{ll}\frac12m\omega_\mathrm{c}^2(x+l_H^2k_y)^2,&x>0,\\
+\infty,&x\leq0,\end{array}\right.
\end{equation}
can be solved (see also Ref.~\cite{Montambaux}) in the quasiclassical approximation using the Einstein-Brillouin-Keller quantization condition \cite{Brack}
\begin{equation}
\int\limits_{x_1}^{x_2}p(x)\,dx=\pi\hbar(n+\alpha),\label{qc_quant1}
\end{equation}
where $E_{nk_y}^\mathrm{qc}$ is the quasiclassical energy level, $p(x)=\{2m[E_{nk_y}^\mathrm{qc}-V(x)]\}^{1/2}$, and the sum of Morse-Maslov indices $\alpha$ equals to $\frac12$ if the potential $V(x)$ is smooth at both turning points $x_{1,2}$, or $\frac34$ if one of these points is located at the hard wall $x=0$.

\begin{figure}[b]
\begin{center}
\includegraphics[width=\columnwidth]{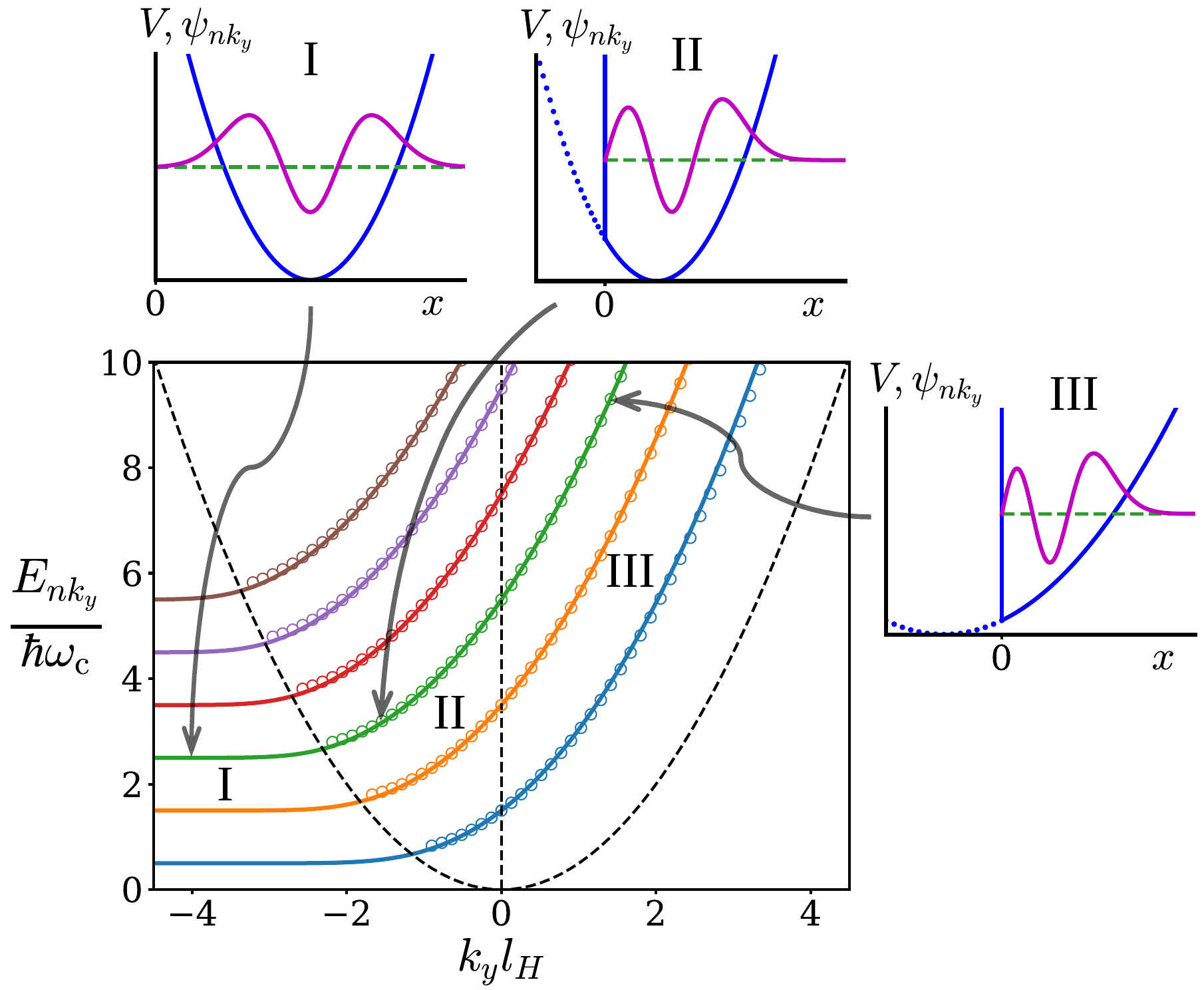}
\end{center}
\caption{\label{Fig_qc}Exact energies of the edge states $E_{nk_y}$ in magnetic field (solid lines) as functions of $k_y$, and their quasiclassical approximations (circles) given by Eq.~(\ref{E_qc_appr}). Insets show quadratic confining potential $V(x)$ and wave function $\psi_{nk_y}(x)$ for $n=2$ in different regions of quasiclassical quantization.}
\end{figure}

Depending on $E_{nk_y}^\mathrm{qc}$ and $k_y$, we identify different regimes of quasiclassical quantization shown in the insets in Fig.~\ref{Fig_qc}. At $k_y<-(2mE_{nk_y}^\mathrm{qc})^{1/2}\hbar^{-1}$ (region I), both turning points lie at the parabolic part of $V(x)$, so we obtain $\int_{x_1}^{x_2}p(x)dx=\pi E_{nk_y}^\mathrm{qc}/\omega_\mathrm{c}$ and hence the familiar harmonic oscillator quasiclassical energies $E_{nk_y}^\mathrm{qc}=\hbar\omega_\mathrm{c}(n+\frac12)$ coinciding with the exact ones. At $-(2mE_{nk_y}^\mathrm{qc})^{1/2}\hbar^{-1}<k_y<0$ (region II) and $k_y>0$ (region III) one of the turning points is the hard wall $x_1=0$, and the quantization condition (\ref{qc_quant1}) reads
\begin{equation}
\frac{r(x)}x=\frac{n+\frac34}\varepsilon,\label{qc_quant2}
\end{equation}
where $x=\varepsilon/z$, $\varepsilon=E_{nk_y}^\mathrm{qc}/\hbar\omega_\mathrm{c}$ is the dimensionless energy, and $z=l_H^2k_y^2/2$. The function
\begin{equation}
r(x)=\frac{x}2-\mathrm{sgn}(k_y)\left(\frac{x}2+\frac{\sqrt{x-1}-x\arctan\sqrt{x-1}}\pi\right)\label{f_exact}
\end{equation}
can be surprisingly well approximated by the simpler expression
\begin{equation}
r(x)\approx\frac{x}2-\mathrm{sgn}(k_y)\sqrt{\frac{4(x-1)}{\pi^2}+\frac14}.\label{f_appr}
\end{equation}
From Eqs.~(\ref{qc_quant2}) and (\ref{f_appr}) we can obtain the approximate quasiclassical energies in the regions II and III:
\begin{align}
E_{nk_y}^\mathrm{qc}&=\frac{4\hbar^2k_y^2}{\pi^2m}+\hbar\omega_\mathrm{c}\left(2n+\frac32\right)\nonumber\\
&+\frac{2\sqrt2\hbar^2k_y}{ml_H}\sqrt{2n+\frac32+\frac{l_H^2k_y^2}2\left(\frac2\pi-\frac\pi4\right)^2},\label{E_qc_appr}
\end{align}
which are quite close to the exact ones as shown in Fig.~\ref{Fig_qc}.

Using the quasiclassical dispersion equation (\ref{qc_quant1}) and the property $xr'(x)-r(x)=\mathrm{sgn}(k_y)\sqrt{x-1}/\pi$ of the exact function (\ref{f_exact}), we obtain the derivative
\begin{equation}
\frac{\partial k_y}{\partial E_{nk_y}^\mathrm{qc}}=\frac\pi{\hbar\omega_\mathrm{c}l_H^2|k_y|}\frac{r'(x)}{\sqrt{x-1}}\label{qc_der}
\end{equation}
taken at $n=\mathrm{const}$. If the equilibrium Fermi level $E_\mathrm{F}$ is located between $n_0$th and $(n_0+1)$th bulk Landau levels, and $n_0\gg1$ (as assumed in the quasiclassical limit), then we can approximate summation over $n$ in Eq.~(\ref{v_dr_harm}) by continuous integration. Using Eq.~(\ref{qc_der}) and taking the quantization condition (\ref{qc_quant2}) at $\varepsilon=\mathrm{const}$ to switch from $n$ to $x$, we obtain
\begin{align}
\sum_{n=0}^{n_0}\frac{\partial k_n^\mathrm{max}}{\partial E_{nk_y}}&\approx\frac\pi{\sqrt2\hbar\omega_\mathrm{c}l_H}\int\limits_0^{n_0}dn\sqrt{\frac{x}\varepsilon}\frac{r'(x)}{\sqrt{x-1}}\nonumber\\
=&\frac1{\sqrt2\hbar\omega_\mathrm{c}l_H}\int\limits_1^\infty\frac{dx\,\sqrt\varepsilon}{x^{3/2}}=\frac1{\hbar\omega_\mathrm{c}l_H}\sqrt{\frac{2E_\mathrm{F}}{\hbar\omega_\mathrm{c}}}.
\end{align}
Substituting this result to Eq.~(\ref{v_dr_harm}) and taking into account that $E_\mathrm{F}\approx\hbar\omega_\mathrm{c}n_0$, we obtain the final quasiclassical expression (\ref{v_dr_qc}) for the drift velocity.

\bibliography{Drift}

\end{document}